# An efficient method for calculating thermoelastic properties

Zhongqing Wu

Department of Chemical Engineering and Materials Science and
Minnesota Supercomputing Institute, University of Minnesota, Minneapolis, MN 55455, USA

First-principles quasi-harmonic calculations play a very important role in mineral physics because they can accurately predict the structure and thermodynamic properties of materials at pressure and temperature conditions that are still challenging for experiments. It also enables calculations of thermoelastic properties by obtaining the second-order derivatives of the free energies with respect to strain. However, these are exceedingly demanding computations requiring thousands of large jobs running on $10^1$ processors each. Here we introduce a simpler approach that requires only calculations of static elastic constants and phonon density of states for unstrained configurations. This approach decreases the computational time by more than one order of magnitude. We show results on MgO and forsterite that are in very good agreement with previous first-principles results and experimental data.

# 1. Introduction

Elasticity is a basic property of materials important in many processes such as brittle failure, flexure and the propagation of elastic waves. The study of the elasticity of the Earth's materials is very important to geophysics. The primary source of information about the Earth's interior is provided by seismological observation. Understanding these observations requires knowledge of the elastic properties of materials at high pressures and temperatures. Although experimental progress has been steady, measuring the elastic constant at the pressure and temperature conditions of the Earth's mantle remains a considerable challenge.

Recently, a complementary approach to experiments [Karki *et al.* 1999] has been developed based on first-principles quasi-harmonic calculations. The approach shows high predictive power and can provide critical information [Karki *et al.* 1999; Wentzcovitch *et al.* 2004; 2006] including that which is difficult to measure experimentally. However, the approach needs to calculate the density of state (DOS) of the phonons for each strain structure. For example, for an orthorhombic crystal, which is a very common structure in minerals, the phonon DOS for at least fifteen strain structures at each volume (repeating the calculation for about 8 volumes to fit the data) is required to get the nine elastic constants. This makes the calculation challenging, especially for crystals with a large primitive cell, which is often the case for the Earth's minerals. Here we report another method that avoids calculating the phonon DOS of strain configurations. Obviously the method can decrease computation time by more than an order of magnitude for most cases, which makes computing efforts the same order as obtaining the isothermal equation of state of materials. We apply our method on MgO and forsterite and the results are in good agreement with previous first-principles results [Karki *et al.* 1999] and experimental data.

## 2. Method

The isothermal elastic constant for the infinitesimal strain [Barron and Klein, 1965] is given by

$$c_{ijkl} = \frac{1}{V}\left(\frac{\partial^2 F}{\partial e_{ij} \partial e_{kl}}\right) + \frac{1}{2}P(2\delta_{ij}\delta_{kl} - \delta_{il}\delta_{jk} - \delta_{ik}\delta_{jl}) \quad (1)$$

Here $F$ is Helmholtz free energy, which is expressed in the quasi-harmonic approximation as

$$F(V,T) = U_0(V) + \frac{1}{2}\sum_{q,m}\hbar\omega_{qm}(V) + k_B T \sum_{q,m}\ln\{1 - \exp[-\hbar\omega_{qm}(V)/k_B T]\} \quad (2)$$

where subscript $q$ is the wave vector and $m$ is index of normal modes. Usually, in order to obtain the derivative of the Helmholtz free energy with respect to all kinds of strains, we need to calculate the phonon DOS not only for a configuration without strain but also for different strain configurations, which number 15 for orthorhombic structures and 5 for cubic ones. All these calculations must be repeated for several volumes (generally >8) to get the pressure and temperature dependence of the elastic constant. Phonon calculation is time-consuming. For example, the computing time for phonon DOS are almost three orders of magnitude greater than that for the self-consistent calculation of MgO. Therefore, calculating the thermal elastic constant by the method described above will necessitate a large computation effort and may even be impracticable for materials with a large primitive cell, which is usually the case for minerals. Obviously, we can save more than one order of magnitude of computation time if we can calculate the thermal elastic constant just from the phonon DOS for configuration without strain.

Phonon DOSs for configuration without strain at several volumes provides information about the volume dependence of phonon frequencies. If we can know how the frequency varies with infinitesimal strain based on its volume dependence, then we can obtain the thermal elastic constant without performing the phonon calculations for strain configuration. Here we derived the relations between frequency dependence on volume and strain for a crystal with orthorhombic symmetry. The relation can also be applied to cubic and tetragonal crystals. The relations will be a little different for crystals with other symmetries; however, the basic idea should be still hold.

*2.1. Relations between volume and strain Grüneisen parameters*

The volume dependence of the frequencies is described by mode Grüneisen parameters as

$$\frac{d\omega_{qm}}{\omega_{qm}} = -\gamma_{qm}\frac{dV}{V} \tag{3}$$

If the volume change is only due to longitudinal strain, we also can show that the frequencies change with

$$\frac{d\omega_{qm}}{\omega_{qm}} = -(\gamma_{1,qm}e_1 + \gamma_{2,qm}e_2 + \gamma_{3,qm}e_3) \tag{4}$$

where, we use the Voigt's notation, $e_1$, $e_2$, and $e_3$ is the longitudinal strain along the three crystal axes, $x_1$, $x_2$ and $x_3$, for orthorhombic crystals, respectively. In order to avoid confusion, we call $\gamma_{1,qm}$, $\gamma_{2,qm}$, and $\gamma_{3,qm}$ strain Grüneisen parameters. We can either express the frequency change by the power of the volume change

$$\begin{aligned}\Delta\omega_{qm} &= \frac{\partial\omega_{qm}}{\partial V}\cdot dV + \frac{1}{2}\frac{\partial^2\omega_{qm}}{\partial V^2}\cdot dV^2 \\ &= -\gamma_{qm}\omega_{qm}[e_1 + e_2 + e_3 - (e_1^2 + e_2^2 + e_3^2)/2] \\ &\quad -\frac{1}{2}V\frac{\partial\gamma_{qm}}{\partial V}\omega_{qm}(e_1 + e_2 + e_3)^2 \\ &\quad +\frac{1}{2}\gamma_{qm}^2\omega_{qm}(e_1 + e_2 + e_3)^2\end{aligned} \tag{5}$$

or express the frequency change by the power of the strain

$$\Delta\omega_{qm} = \sum_i \frac{\partial \omega_{qm}}{\partial e_i} \cdot e_i + \frac{1}{2}\sum_{i,j} \frac{\partial^2 \omega_{qm}}{\partial e_i \partial e_j} \cdot e_i e_j$$

$$= -\omega_{qm}(\gamma_{1,qm}e_1 + \gamma_{2,qm}e_2 + \gamma_{3,qm}e_3 - \frac{1}{2}\gamma_{1,qm}e_1^2 - \frac{1}{2}\gamma_{2,qm}e_2^2 - \frac{1}{2}\gamma_{3,qm}e_3^2) \quad (6)$$

$$-\frac{1}{2}\omega_{qm}(\frac{\partial \gamma_{1,qm}}{\partial e_1}e_1^2 + \frac{\partial \gamma_{2,qm}}{\partial e_2}e_2^2 + \frac{\partial \gamma_{3,qm}}{\partial e_3}e_3^2 + \frac{2\partial \gamma_{1,qm}}{\partial e_2}e_1 e_2 + \frac{2\partial \gamma_{2,qm}}{\partial e_3}e_2 e_3 + \frac{2\partial \gamma_{3,qm}}{\partial e_1}e_3 e_1)$$

$$+\frac{1}{2}\omega_{qm}[(\gamma_{1,qm})^2 e_1^2 + (\gamma_{2,qm})^2 e_2^2 + (\gamma_{3,qm})^2 e_3^2 + 2\gamma_{1,qm}\gamma_{2,qm}e_1 e_2 + 2\gamma_{2,qm}\gamma_{3,qm}e_2 e_3 + 2\gamma_{3,qm}\gamma_{1,qm}e_3 e_1]$$

For any longitudinal strain, both equations should have the same frequency change. According to the fist-order term of two equations, we get one of relationships

$$\gamma_{1,qm}\frac{e_1}{(e_1+e_2+e_3)} + \gamma_{2,qm}\frac{e_2}{(e_1+e_2+e_3)} + \gamma_{3,qm}\frac{e_3}{(e_1+e_2+e_3)} = \gamma_{qm} \quad (7\text{-}1)$$

This relation also ensures that the last term in both equations are equal each other. In order to keep the second term in both equations equal for any longitudinal strain, the corresponding coefficients in both equation should have relation as following

$$\frac{\frac{\partial \gamma_{i,qm}}{\partial e_j}}{\gamma_{i,qm}\gamma_{j,qm}} = \frac{V\frac{\partial \gamma_{qm}}{\partial V}}{\gamma_{qm}^2} \quad (7\text{-}2)$$

Therefore we can introduce two azimuth parameters $\theta_{qm}$ and $\phi_{qm}$, which we call the Grüneisen azimuth, and rewrite the equation (7-1)

$$\gamma_{1,qm} = \frac{e_1+e_2+e_3}{e_3}\gamma_{qm}\cos^2\theta_{qm}\sin^2\phi_{qm}$$

$$\gamma_{2,qm} = \frac{e_1+e_2+e_3}{e_2}\gamma_{qm}\sin^2\theta_{qm}\sin^2\phi_{qm} \quad (8)$$

$$\gamma_{3,qm} = \frac{e_1+e_2+e_3}{e_3}\gamma_{qm}\cos^2\phi_{qm}$$

Each mode m and wave vector q corresponded to a Grüneisen azimuth. All these Grüneisen azimuth will form a distribution function $f(\theta,\phi)$. What we need in calculating elastic constant is the average value about strain Grüneisen parameters, namely something like

$$\overline{\gamma_{i,qm}} = \frac{1}{3N}\sum_{q,m} w_q \gamma_{i,qm} = \int \gamma_{i,qm}(\theta,\phi) f(\theta,\phi) \sin\phi d\theta d\phi \qquad (9)$$

where, $w_q$ is the weight at wave vector $q$ and $3N$ are the number of normal modes. The distribution function $f(\theta,\phi)$ is unknown for most materials except one case, isotropic materials. For isotropic materials, $f(\theta,\phi)$ is a constant and should be equal to $1/4\pi$. The symmetry reduction seems not change the distribution function $f(\theta,\phi)$ significantly. For example, the tetragonal system, produced by applying a small strain in a cubic system, and its original cubic system should have almost the same distribution function $f(\theta,\phi)$ since the small strain does not change the strain Grüneisen parameter, which is reflected by equation (4). Therefore we assume that the uniform distribution function $f(\theta,\phi)$ is a good approximation for the crystal.

Then we can calculate all kinds of average values of strain Grüneisen parameters, which are needed in calculating the thermal elastic constant.

$$\begin{aligned}\overline{\gamma_{i,qm}} &= \frac{e_1+e_2+e_3}{e_i}\frac{1}{4\pi}\iint \gamma_{qm} \cos^2\theta_{qm} \sin^2\phi_{qm} d\Omega \\ &\simeq \frac{e_1+e_2+e_3}{e_i}\frac{\overline{\gamma_{qm}}}{4\pi}\iint \cos^2\theta_{qm} \sin^2\phi_{qm} d\Omega \\ &= \frac{e_1+e_2+e_3}{3e_i}\overline{\gamma_{qm}}\end{aligned} \qquad (10)$$

Similarly, we have

$$\overline{\gamma_{i,qm}\gamma_{j,qm}} = \begin{cases} \dfrac{1}{5}\dfrac{(e_1+e_2+e_3)^2}{e_i e_j}\overline{(\gamma_{qm})^2} & \text{if } i=j \\ \dfrac{1}{15}\dfrac{(e_1+e_2+e_3)^2}{e_i e_j}\overline{(\gamma_{qm})^2} & \text{if } i\neq j \end{cases} \quad (11)$$

By using the relation 2 (Eq. (7-2)), we can get

$$\overline{\dfrac{\partial \gamma_{i,qm}}{\partial e_j}} = \begin{cases} \dfrac{1}{5}\dfrac{(e_1+e_2+e_3)^2}{e_i e_j}V\overline{\dfrac{\partial \gamma_{qm}}{\partial V}} & \text{if } i=j \\ \dfrac{1}{15}\dfrac{(e_1+e_2+e_3)^2}{e_i e_j}V\overline{\dfrac{\partial \gamma_{qm}}{\partial V}} & \text{if } i\neq j \end{cases} \quad (12)$$

where $i,j = 1, 2, 3$. Although we assume the isotropic distribution function, the average value about the strain Grüneisen parameters are anisotropy, which are reflected by the axis compressive ratio under pressure, namely $e_1:e_2:e_3$. We know the axis compressive ratio $e_1:e_2:e_3$ at different pressures from first-principles static calculations. For cubic $e_1 = e_2 = e_3$. For other crystals, they are not equal in general and change with volume but are known. The volume dependence of $\gamma_{qm}$ is known since we had calculated the frequencies at different pressures (or volume). Therefore, we can calculate all kinds of average values about strain Grüneisen parameters using the above equation.

*2.2. Thermal elastic constant*

*2.2.1. Isothermal longitudinal and off-diagonal elastic constant*

For orthorhombic crystals, there are nine elastic constants. Using Voigt's notation, the isothermal elastic constant in equation (1) excluding the shear elastic constant can be rewritten as

$$c_{ij}(V,T) = \dfrac{1}{V}\left(\dfrac{\partial^2 F(V,T)}{\partial e_i \partial e_j}\right) + (1-\delta_{ij})P(V,T) = c_{ij0}(V) + c_{ijPH}(V,T) \quad (13)$$

where $i,j = 1, 2, 3$. The first term $c_{ij0}(V) = \frac{\partial^2 U}{\partial e_i \partial e_j} + (1-\delta_{i,j})P_0(V)$ is the elastic constant from the local-density approximation (LDA) static calculation, where $P_0(V)$ is the pressure from the LDA static calculation. All phonon contributions to the elastic constant are included in the second term, which can be written as

$$c_{ijPH}(V,T) = c_{ij\omega}(V) + c_{ijT}(V,T) + (1-\delta_{ij})P_T(V,T) \qquad (14)$$

where $P_T(V,T) = P(V,T) - P_0(V)$ is the pressure resulting from the phonon contribution. Zero motion and temperature contributions to the elastic constant are described by the first and second terms of equation (14), respectively,

$$c_{ij\omega}(V) = \frac{\hbar}{2V} \sum_{qm} \frac{\partial^2 \omega_{qm}(V)}{\partial e_i \partial e_j} = \frac{3\hbar N}{2V}(\overline{\gamma_{i,qm}\gamma_{j,qm}} - \overline{\frac{\partial \gamma_{i,qm}}{\partial e_j}} + \delta_{i,j}\overline{\gamma_{i,qm}}) \qquad (15)$$

$$c_{ijT}(V,T) = \frac{k_B T}{V} \sum_{q,m} \frac{\partial^2 \ln(1-e^{-Q_{qm}})}{\partial e_i \partial e_j} \qquad (16)$$

$$= \frac{k_B T}{V} \sum_{q,m} [-Q_{qm}^2 \cdot \frac{e^{Q_{qm}}}{(e^{Q_{qm}}-1)^2} \cdot \gamma_{i,qm}\gamma_{j,qm} + Q_{qm} \cdot \frac{1}{e^{Q_{qm}}-1} \cdot (\gamma_{i,qm}\gamma_{j,qm} - \frac{\partial \gamma_{i,qm}}{\partial e_j} + \gamma_{i,qm}\delta_{ij})]$$

where $Q_{qm} = \hbar\omega_{qm}/k_B T$ and 3N is the number of normal modes. $Q_{qm}^2 e^{Q_{qm}}/(e^{Q_{qm}}-1)^2$ and $Q_{qm}/(e^{Q_{qm}}-1)$ are weakly frequency dependent especially at high temperature. Therefore, it is a good approximation to assume that $Q_{qm}^2 e^{Q_{qm}}/(e^{Q_{qm}}-1)^2$ and $Q_{qm}/(e^{Q_{qm}}-1)$ do not correlate with $\gamma_{i,qm}$. Then we have

$$c_{ijT}(V,T) \simeq \frac{3k_B TN}{V}[-\overline{Q_{qm}^2 \cdot \frac{e^{Q_{qm}}}{(e^{Q_{qm}}-1)^2}} \cdot \overline{\gamma_{i,qm}\gamma_{j,qm}} + \overline{Q_{qm} \cdot \frac{1}{e^{Q_{qm}}-1}} \cdot (\overline{\gamma_{i,qm}\gamma_{j,qm}} - \overline{\frac{\partial \gamma_{i,qm}}{\partial e_j}} + \overline{\gamma_{i,qm}}\delta_{ij})] \qquad (17)$$

*2.2.2. The adiabatic elastic constant*

The relation between the adiabatic and isothermal elastic constant is described [Davies 1974] by

$$c_{ij}^S(V,T) = c_{ij}(V,T) + \frac{VT\lambda_i\lambda_j}{C_V}, \tag{18}$$

where $i,j = 1, 2, 3$ and $C_V$ is heat capacity at constant volume, and

$$\lambda_i = \frac{1}{V}\frac{\partial S(V,T)}{\partial e_i} = \frac{k_B}{V}\sum_{q,m} Q_{qm}^2 \frac{e^{Q_{qm}}}{(e^{Q_{qm}}-1)^2}\gamma_{i,qm} = \frac{3k_B N}{V}\overline{Q_{qm}^2 \frac{e^{Q_{qm}}}{(e^{Q_{qm}}-1)^2}\gamma_{i,qm}} \tag{19}$$

*2.2.3. Shear elastic constant*

Similarly, the shear elastic constant can also be expressed as

$$c_{kk}(V,T) = c_{kk0}(V) + c_{kkPh}(V,T), \tag{20}$$

where $k = 4, 5, 6$. Up to now, we still do not know how to calculate the second term of the equation, the phonon contribution to the shear elastic constant, because we only discuss the longitudinal strain. However, as Figure 1 shows, the shear strain in the $x_1$ and $x_2$ planes can be regarded as a combined strain with a compressive strain along the $x_{1'}$ axis and a tensile strain along the $x_{2'}$ axis. Energy change are be described by

$$\frac{\Delta E}{V} = \frac{1}{2}c_{ijkl}e_{ij}e_{kl} \tag{21}$$

If volume is conserve in strain. In X1 and X2 coordinate, it is

$$\begin{aligned}\frac{\Delta E}{V} &= \frac{1}{2}(c_{1212}e_6e_6 + c_{1221}e_6e_6 + c_{2112}e_6e_6 + c_{2121}e_6e_6)\\&= 2c_{66}e_6e_6\end{aligned} \tag{22}$$

In X1' and X2' coordinate, it is

$$\frac{\Delta E}{V} = \frac{1}{2}(c_{1'1'1'1'}e_{1'}e_{1'} + c_{1'1'2'2'}e_{1'}e_{2'} + c_{2'2'1'1'}e_{2'}e_{1'} + c_{2'2'2'2'}e_{2'}e_{2'})$$
$$= \frac{1}{2}(c_{1'1'}e_{1'}e_{1'} + c_{1'2'}e_{1'}e_{2'} + c_{2'1'}e_{2'}e_{1'} + c_{2'2'}e_{2'}e_{2'}) \tag{23}$$

The strain relation in two coordinates is

$$e_{1'} = e_6, \quad e_{2'} = -e_6 \tag{24}$$

With Eq. (21) –(24), we got

$$c_{66Ph} = (c_{1'1'Ph} + c_{2'2'Ph} - 2c_{1'2'Ph})/4 \tag{25-1}$$

The equation (25) builds the relation between the shear elastic constant and longitudinal and off-diagonal elastic constant, which we already know hot to calculate. The formula used to calculated $c_{ijPh}$ can be applied to $c_{i'j'Ph}$, here i,j=1.3. The only difference is that $e_i \gamma_{qm}$ should be replaced by $e_{i'} \gamma'_{qm}$ The strain in $X_1$, $X_2$ and $X_3$ $e_i$ (i=1,3) transformed into the stain in $X_{1'}$, $X_{2'}$ and $X_3$ as fellows:

$$e_{1'} = \frac{e_1 + e_2}{2}; \quad e_{2'} = \frac{e_1 + e_2}{2}; \quad e_{3'} = e_3; \quad e_{6'} = e_1 - e_2 \tag{26}$$

When we use the equations in section 2.2.1. to calculate the $c_{i'j'Ph}$ by replacing $e_1, e_2$ and $e_3$ with $e_{1'}, e_{2'}$ and $e_{3'}$, in principle we need new volume Grüneisen parameters $\gamma'_{q,m}$ corresponding to strain $e_{1'} = \frac{e_1 + e_2}{2}; e_{2'} = \frac{e_1 + e_2}{2}; e_{3'} = e_3; e_{6'} = 0$. $\gamma'_{q,m}$ are different from $\gamma_{q,m}$ which correspond to strain $e_{1'} = \frac{e_1 + e_2}{2}; e_{2'} = \frac{e_1 + e_2}{2}; e_{3'} = e_3; e_{6'} = e_1 - e_2$ because non zero $e_{6'}$ Fortunately, the difference between $\gamma'_{q,m}$ and $\gamma_{q,m}$ should be small because (i) in contrast to the compressive strain, the shear strain $e_{6'}$ should be relatively small as shown in equation (26); (ii) the frequency changes caused by the shear strain are far smaller than the frequency changes caused by the compressive strain hence contribution to Grüneisen parameters from the shear strain are relatively small. Therefore we can still use $\gamma_{q,m}$ to calculate the shear elastic constant without sacrificing too much precision. $c_{44}$ and $c_{55}$ can be calculated similarly.

## 3. Results and discussion

Our results for the thermal elasticity in two major minerals, cubic MgO (periclase) and orthorhombic forsterite (α phase of $Mg_2SiO_4$), are presented here. The calculations are performed using the density functional theory with the local-density approximation [Perdew and Zunger, 1981]. Calculation details are similar to those reported in previous works [Karki *et al.* 1999; Li *et al.* 2007]. Brillouin-zone sampling for electronic (phonons) states was carried out on 10 (8) and 2 (5) special *k*-points (*q*-points) for periclase and forsterite, respectively. Phonon frequencies were calculated using density functional perturbation theory [Baroni 2001].

As shown in Figure 2, our calculated elastic moduli of MgO at zero pressure are almost the same as our previous calculations [Kariki *et al.* 2000], which require the phonon DOS of strain configurations, except there is some discrepancy for $c_{44}$ above 2000 K and $c_{12}$ above 1200 -K. Both results agree well with the experimental data [Isaak *et al.*, 1989a] for a very wide temperature range. At temperatures above ~1000 K, the $c_{11}$ appear to deviate more sharply from the experimental trend. The deviations result from overestimating the thermal expansivity by using the quasi-harmonic approximation (QHA) at those temperatures [Kariki *et al.* 2000]. Unexpectedly, our results seem more consistent with the experiment data than our previous results, especially the slope of $c_{44}$, although our new method adopts the approximation. The possible reason is that our method avoids the direct calculation of the second derivative of the free energy with respect to infinitesimal strain, which is sensitive to fitting conditions such as which equation of state is adopted and how many points are used in fitting. In contrast, our results are insensitive to fitting conditions. Furthermore the approximation should work very well for cubic crystals.

The elastic moduli of forsterite at zero pressure also are consistent with the experimental data (see Fig. 3). At room temperature, the longitudinal elastic constants $c_{11}$, $c_{22}$ and $c_{33}$ are slightly smaller than the experimental data [Issak *et al.*, 1989b]. However, their temperature derivatives are in excellent agreement with the experimental results. At temperatures above ~1000 K, the longitudinal elastic constant of forsterite decreases with temperature more slowly than the experimental data. This similar temperature behavior also appears in the bulk modulus, $K$s, (see Fig. 3). The bulk moduli are calculated without using any of the approximations mentioned above, the deviation from experimental data should also be attributed to the anharmonic effect as in the case of MgO. However, MgO and forsterite show completely different anharmonic behavior. The longitudinal elastic constant of MgO decreases more quickly than the experimental data. On the contrary, the longitude elastic constant of forsterite decrease more slowly than the experimental data. This behavior is consistent with the thermodynamic observation of anharmonicity. The thermal expansivity is overestimated in MgO but underestimated in forsterite by the QHA [Li *et al.* 2007]. The heat capacity $C_v$ of forsterite exceeds the Dulong and Petit limit at about 1400 K [Gillet 1991, Bouhifd 1996 , Anderson 1996)]. In contrast, the heat capacity $C_v$ of MgO is predicted to be 2.6% smaller than the Dulong and Petit limit even at T = 2000 K [Anderson and Zou, 1990].

Our calculations can also reproduce high-pressure experimental data well. Figure 4 shows the dependence of velocity of on pressure at various temperatures for forsterite. Our first principles static calculation is in agreement with previous results by da Silva *et al.* [1997]. Both calculations overestimate the compressive velocity. After including zero motion and the temperature contributions in the elastic constant, the velocity becomes consistent with the experiment observations [Zha *et al.* 1996; Li *et al.* 2005; Issak *et al.* 1989] at both 300 K and 1070 K. The temperature contribution to the velocity decreases with increasing the pressure from 0.47 (0.30) m/s/K at 0 GPa to 0.32 (0.19) m/s/K at 10 GPa for the compressive (shear) velocity.

## 4. Conclusion

We derived the relation between the strain Grüneisen parameters and volume Grüneisen parameters, from which we introduce a Grüneisen azimuths for each phonon mode. By assuming the Grüneisen azimuths have a homogenous distribution, we use the formulas to calculate all kinds of average values of the strain Grüneisen parameters just requiring the axis compressive ratio $e_1:e_2:e_3$ and volume Grüneisen parameters $\gamma_{qm}$ at different pressures, which can be obtained from first-principles static calculations and phonon DOS of unstrain configurations. Then we can calculate the thermal elastic constant without requiring the phonon DOS of the strain configuration, which reduces the computing time by more than an order of magnitude compared with the previous method. This kind of saving in computing time is very useful in investigating the elasticity of minerals because the phonon DOS calculation is very demanding and usually becomes too unwealdy for minerals because of their large primitive cell size. The calculated elasticity on periclase and forsterite shows that our method includes the phonon contribution well both at zero pressure and at high pressure and works well both for cubic and orthorhombic crystals.

**Acknowledgements**

Calculations were performed with the Quantum ESPRESSO package (http://www.pwscf.org). This work was supported by the National Science Foundation under grant numbers NSF/EAR 0230319, 0635990, and NSF/ITR 0428774 (VLab).

**Figure Captions**

**Figure 1.** The strain relations in different coordinate axis. The structure before and after strain are denoted by green and red lines, respectively. The structure experiences a shear strain as shown by the solid line if viewing from the $x_1$ and $x_2$ axis. This shear strain can also be viewed as the combined strain with a compressive strain along the $x_{1'}$ axis and a tensile strain along the $x_{2'}$ axis as shown by the dashed lines.

**Figure 2.** The temperature dependence of the elastic constant of MgO at zero pressure (solid line), compared to previous results [Karki *et al.* 1999] denoted by the dashed line and experimental data [Issak *et al.* 1989a] denoted by open circles.

**Figure 3.** The temperature dependence of the elastic modulus of forsterite at zero pressure (solid line), compared to experimental data [Issak *et al.* 1989b] denoted by solid squares and circles.

**Figure 4.** The pressure dependence of the velocity of forsterite at various temperatures, compared with experiment data from Issak *et al.* [1989b] (stars), Zha *et al.*[1996] (open circles) and Li *et al.* [2003] (solid circles).

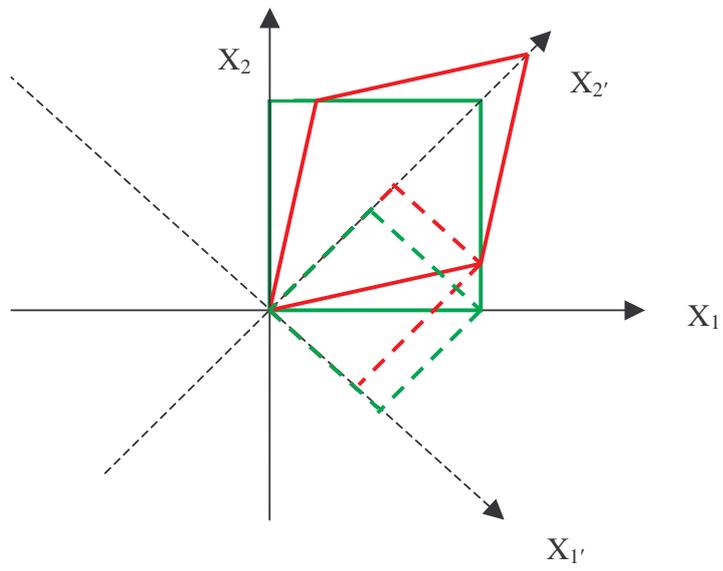

Fig .1

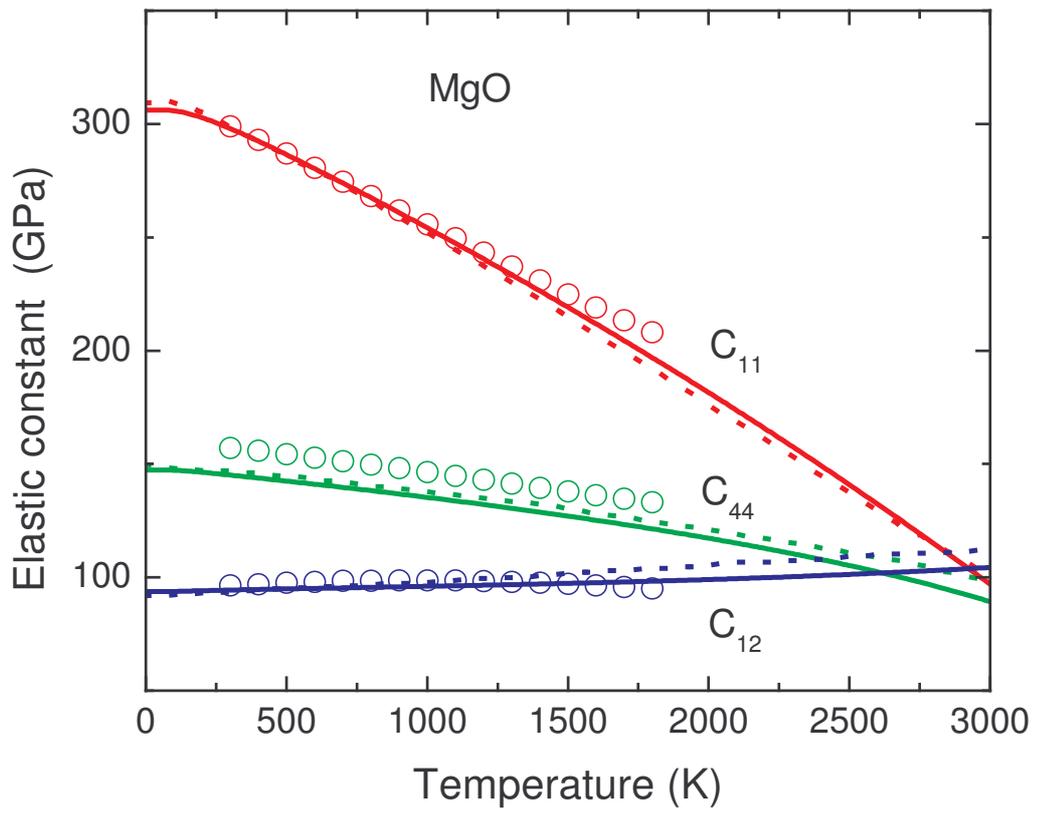

Fig. 2

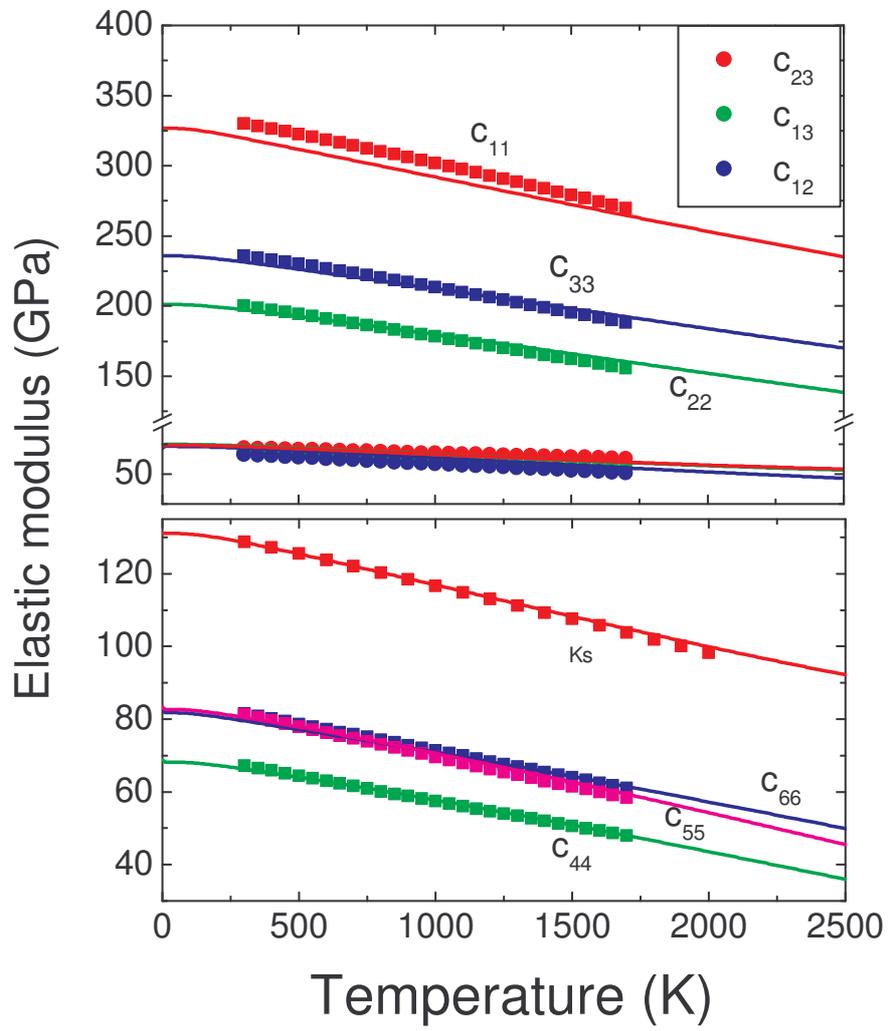

Fig. 3

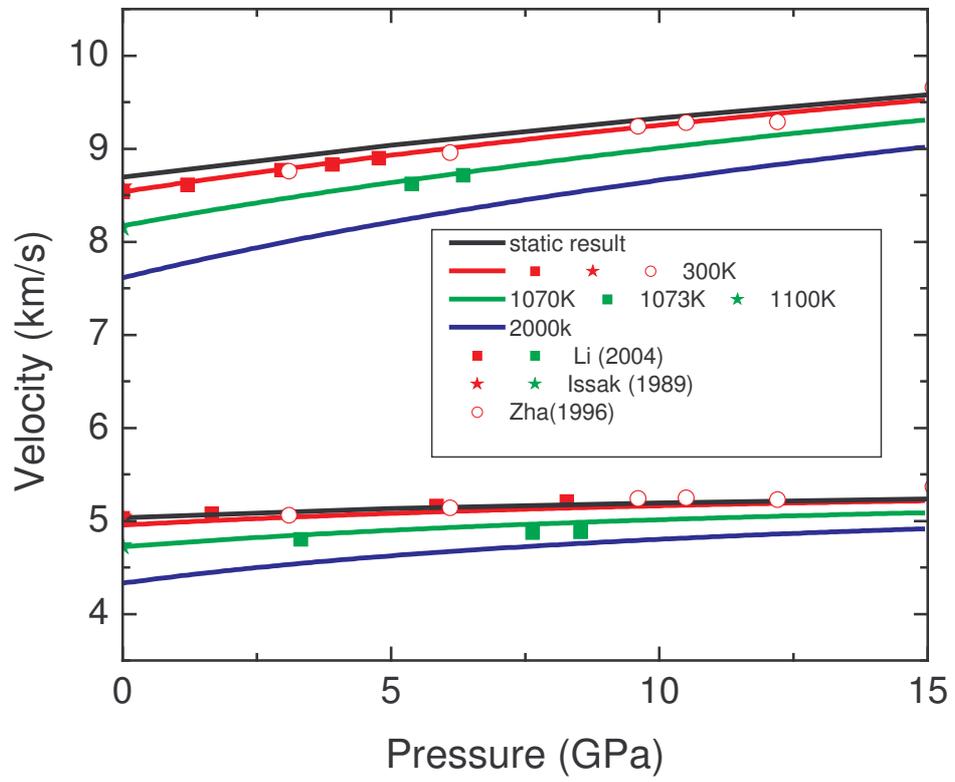

Fig. 4